\documentclass[aps,twocolumn,showpacs,eqsecnum,amsmath,amssymb]{revtex4}
\usepackage{graphicx}
\usepackage{dcolumn}
\usepackage{bm}
\begin{document}

\title{Thermal effect on primordial black holes in standard Higgs minimum double-well potential}
\author{Xi-Bin Li}
\email{lxbbnu@mail.bnu.edu.cn}
\affiliation{Department of Physics, Beijing Normal University, Beijing 100875, China}
\affiliation{School of Physics and Technology, Wuhan University, 430072 Wuhan, China}
\author{Jian-Yang Zhu}
\thanks{Corresponding author}
\email{zhujy@bnu.edu.cn}
\affiliation{Department of Physics, Beijing Normal University, Beijing 100875, China}

\date{\today}

\begin{abstract}
We attempt a new scheme to combine the Higgs field in the minimal standard model and the statistic physics with thermal effect together. By introducing the stochastic differential equation in FRW metric frame which is something like the warm inflation model but not exactly the same. By using the previous researches on Fokker-Planck equation with double-well potential, we find the abundance of primordial black holes (PBHs) dominate at a special mass and the PBHs with extremely large or extremely small mass could be almost excluded. In addition, two perturbed model within this frame are employed, one is the model with symmetry breaking and another is stochastic resonance. The former may increase the probability to the generation of PBHs, while the latter may both increase and decrease the probability. Finally, we also discuss the possibility on extension this scenario to other models.
\end{abstract}
\pacs{98.80.-k, 98.80.Bp, 98.80.Es, 05.70.Ce}

\maketitle

\section{\label{introduction}Introduction}
Cosmological inflation \cite{STAROBINSKY198099,Sato1980yn,PhysRevD.23.347,LINDE1983177,Linde1981mu} is a period of exponentially expansion driven by very high vacuum energy at the early Universe, during which the perturbations outside horizon freeze out to become the primordial cosmological perturbations that leads to the isotropies of cosmic microwave background and the large-scale structure \cite{Mukhanov1982nu,PhysRevLett.49.1110,PhysRevD.28.679}. Observations on cosmic microwave background constrain on the primordial perturbation $\zeta$ to ba small with level $\zeta\simeq 10^{-5}$ on large scale. The small scale, however, may be sufficiently large to the level of unit. With this level of perturbation, the gravity overcomes the pressure force and collapse to primordial black holes \cite{Hawking1971ei,Carr1974nx,1975ApJ...201....1C}.

The formation of primordial black holes during inflation is achieved by chaotic new inflation which is described by the Fokker-Planck equation \cite{PhysRevD.57.7145,PhysRevD.58.083510}. One of the most widely studied potential is the one with double-well which is used to explain the symmetry breaking in gauge field theory. The one-loop finite-temperature effective potential of the Higgs field in the minimal standard model is well approximated by \cite{PhysRevD.46.550,PhysRevD.45.2685,SHER1989273}
\begin{eqnarray}
    V[\phi]=\frac{1}{2}D(T^2-T_C^2)\phi^2-\frac{1}{2}ET\phi^3+\frac{1}{4}\lambda\phi^4, \label{double_well_potential}
\end{eqnarray}
where $T_C$ denotes the critical temperature. If $T<T_C$, the formula (\ref{double_well_potential}) represents a potential with double wells distributing at both sides of $\phi=0$. The previous work on this model only fucus on the process of quantum tunnelling during which bubble nucleation forms the primordial black holes \cite{RevModPhys.57.1}.
This method, however, neglects an important effect, called thermal effect, because of the existent of temperature.

Except the ignorance of the thermal effect, another problem is that the abundance of primordial black holes should be the same with different masses because the quantum fluctuation $f(t)$ is a Gaussian noise in chaotic inflationary model \cite{LINDE1983177,Pattison2017mbe,Zaballa_2007}
\begin{eqnarray}
    \ddot{\phi}+3H\dot{\phi}+V''[\phi]=D^{1/2}f(t). \label{chaotic_inflation}
\end{eqnarray}

To explain these two problems, we need to extend our sights to search more powerful scenarios. One present and widely studied scenario is the frame within warm inflation which also follows a stochastic differential equation:
\begin{eqnarray}
    \frac{\partial^2}{\partial t^2}\phi(\textbf{x},t)+[3H+\Gamma]\frac{\partial}{\partial t}\phi(\textbf{x},t)&-&\frac{1}{a^2(t)}\nabla^2\phi(\textbf{x},t)\nonumber \\
    +V'[\phi] &=& \xi(\textbf{x},t), \label{SDE_Sec1}
\end{eqnarray}
where the dissipative coefficient $\Gamma$ and fluctuational noise $\xi$ follow the dissipation-fluctuation relation \cite{PhysRevD.76.083520,PhysRevD.91.083540, PhysRevD.84.103503,PhysRevLett.117.151301,1475-7516-2011-09-033, PhysRevD.53.5437, PhysRevD.54.7181, MUKHANOV199852, PhysRevD.59.123512, STAROBINSKY2001383, PhysRevD.64.083514, 0264-9381-21-2-002, PhysRevD.96.103533}. The potential $V[\phi]$ is just the one defined in Eq.~(\ref{double_well_potential}).

With this scenario, stochastic dynamics with double-well potential in thermodynamics is introduced within the frame of Browns' motion in Sec.~\ref{double_well} that is widely used in multiple subjects \cite{KRAMERS1940284,doi:10.1063/1.436049,DYKMAN198553,PhysRevC.67.064606} and deeply studied during past decades. Based on the fundamentally theory of this model, we calculate the abundance of primordial black holes in Sec.~\ref{primordial_black_holes} and numerical results are give in Sec.~\ref{Numerical}. Then, in Sec.~\ref{perturbation}, we calculate the abundances of primordial black holes in symmetry breaking model and stochastic resonance model. In Sec.~\ref{conclution_discussion}, finally, we have a brief conclusion to this paper and some further discussions are given as well which may shed some light into the study of some other models.

\section{\label{double_well}A brief introduction to stochastic dynamics with double-well potential}          
To be better understanding the calculations in next sections, it's necessary to have a introduction to the stochastic dynamical properties of a particle moved in a double-well potential. In this section, we only list the relevant results or properties of such stochastic equation, but the references are given.

\begin{figure}
  \centering
  \includegraphics[width=3.0in,height=3in]{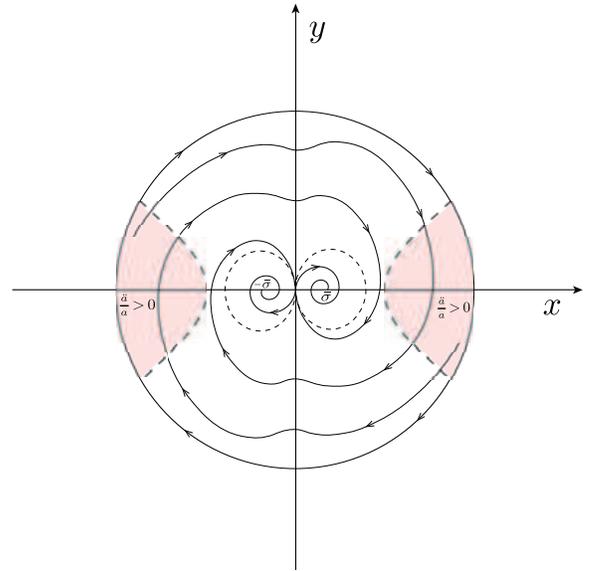}
  \caption{Global phase portrait of dynamic system~(\ref{Langevin_inertial}).}\label{Hopf_bifurcation}
\end{figure}

The potential with double-well reads
\begin{eqnarray}
    V(x)=\frac{1}{2}ax^2+\frac{1}{4}bx^4, \label{double_well_Sec2}
\end{eqnarray}
with $a<0$. The stochastic dynamical differential equation for the inertial translational Brownian motion of a particle of mass $m$ in the potential (\ref{double_well_Sec2}) is given by \cite{BLOMBERG197749}
\begin{eqnarray}
    m\ddot{x}(t)+\zeta\dot{x}(t)+ax(t)+bx^3(t)=F(t), \label{Langevin_inertial}
\end{eqnarray}
where $\zeta$ is the viscous drag coefficient, and $F(t)$ is the Gaussian noise. As discussed in Ref.~\cite{PhysRevD.98.043510}, the singularity in phase space $(x,\dot{x})=(0,0)$ is a Hopf bifurcation point \cite{chow1994normal,chow2012methods} and its global dynamical phase portrait is plotted in Fig. \ref{Hopf_bifurcation}.

In the overdamped limit or non-inertial approximation, where the inertial term $m\ddot{x}$ may be neglected, the stochastic differential equation (\ref{Langevin_inertial}) becomes
\begin{eqnarray}
    \zeta\dot{x}(t)+ax(t)+bx^3(t)=F(t). \label{Langevin_noninertial}
\end{eqnarray}
The dissipative coefficient $\zeta$ and fluctuation noise $F(t)$ follow the dissipation-fluctuation relation \cite{coffey2012langevin}
\begin{eqnarray}
    \langle F(t)F^*(t')\rangle=2 k_\text{B}T\zeta\delta(t-t'). \label{dissipative_fluctuation_Sec2}
\end{eqnarray}
The static or equilibrium solution to Eq. (\ref{Langevin_noninertial}) is
\begin{eqnarray}
    W_0(x)=Z^{-1}\text{e}^{-Ax^2-Bx^4}, \label{Equilibrium_Sec2}
\end{eqnarray}
with $A=a/2k_\text{B}T$, $B=b/4k_\text{B}T$, and
\begin{eqnarray}
    Z=\int_{-\infty}^{+\infty}{\text{d}x\text{e}^{-Ax^2-Bx^4}}. \label{Z_Sec2}
\end{eqnarray}
The variance within static state is
\begin{eqnarray}
    \langle x^2\rangle_0&=& Z^{-1}\int_{-\infty}^{+\infty}{\text{d}xx^2\text{e}^{-Ax^2-Bx^4}} \nonumber\\
    &=& \frac{1}{2\sqrt{2\pi B}}\frac{D_{-\frac{3}{2}}(z)}{D_{-\frac{1}{2}}(z)}, \label{variance_Sec2}
\end{eqnarray}
with $z=A/\sqrt{2B}$ and $D_\nu(z)$ denoting the parabolic cylinder function of the $\nu$th order. The Fokker-Planck equation for the probability distribution function $W$ derived from Eq.(\ref{Langevin_noninertial}) reads \cite{risken2012fokker}
\begin{eqnarray}
    \zeta\frac{\partial}{\partial t}W&=&\mathcal{L}_\text{FP}W \nonumber\\
    &=&\frac{\partial}{\partial x}\left(W\frac{\partial}{\partial x}V\right)+k_\text{B}T\frac{\partial^2}{\partial^2 x}W, \label{Fokker_Planck_Sec2}
\end{eqnarray}
with initial condition $W_2(x,t_0;x_0,t_0)=\delta(x-x_0)$ and Fokker-Planck operator $\mathcal{L}_\text{FP}$. The solution to Eq.(\ref{Fokker_Planck_Sec2}) is called conditioned autocorrelation function.

The most important characteristic variable is the autocorrelation function \cite{doi:10.1063/1.472079,doi:10.1063/1.464598,KALMYKOV2007412}, which is defined as
\begin{eqnarray}
    & &C(t)=\langle x(t)x_0\rangle \nonumber\\
    &=&\int_{-\infty}^{+\infty}\int_{-\infty}^{+\infty}xx_0W_2(x,t|x_0,0)W_0(x_0)\text{d}x\text{d}x_0 \nonumber\\
    &\simeq&\langle x^2\rangle_0\left[\Delta_1\text{e}^{-\lambda_1t}+(1-\Delta_1)\text{e}^{-t/\tau_W}\right]. \label{autocorrelation_Sec2}
\end{eqnarray}
In the equations above, $\lambda_1$ is the minimum non-vanishing eigenvalue of Fokker-Planck operator $\mathcal{L}_\text{FP}$, which is just double times of Kramers' escape rate $r_K$ \cite{risken2012fokker,doi:10.1063/1.2140281,EDHOLM1979313}
\begin{eqnarray}
    r_k=\frac{\sqrt{|V''(0)V''(x_\text{min})|}}{2\pi \zeta}\exp{\left[-\frac{V(0)-V(x_\text{min})}{k_\text{B}T}\right]}, \label{rK_Sec2}
\end{eqnarray}
where $x_\text{min}$ is the minimum of the well. Kramers' escape rate $r_K$ describes a particle's escape rate from one well into another in unit time.
$\Delta_1$ and $\tau_W$ in Eq.(\ref{autocorrelation_Sec2}) are two parameters which also involve two important characteristic times, global time $\tau_\text{int}$ and effective time $\tau_\text{eff}$. The global time is defined as
\begin{eqnarray}
    & &\tau_\text{int}\equiv \frac{\int_0^\infty C(t)\text{d}t}{C(0)} \nonumber\\
    &=&\frac{\tau_0\text{e}^{z^2/4}}{\sqrt{\pi}D_{-\frac{3}{2}}(z)}\sum_{n=0}^\infty{(-1)^n\Gamma(n+\frac{1}{2})D_{-n-\frac{3}{2}}(z)D_{-n-1}(z)} \nonumber\\
    &\simeq&\tau_0 \frac{\text{e}^q-1}{2q}\left(\pi\sqrt{q}+2^{1-\sqrt{q}}\right), \label{Tint_Sec2}
\end{eqnarray}
where $q=z^2/2$ and $\tau_0$ is the characteristic relaxation time defined as
\begin{eqnarray}
    \tau_0=\frac{\zeta}{\sqrt{2bk_\text{k}T}}. \label{relaxation_time_Sec2}
\end{eqnarray}
While the effective time reads
\begin{eqnarray}
    \tau_\text{eff}=\tau_0\frac{D_{-\frac{3}{2}}(z)}{D_{-\frac{1}{2}}(z)}. \label{effective_time_Sec2}
\end{eqnarray}
Thus, we can write the expressions of $\Delta_1$ and $\tau_W$ respectively \cite{coffey2012langevin}
\begin{eqnarray}
    \Delta_1=\frac{\tau_\text{int}/\tau_\text{eff}-1}{\lambda_1\tau_\text{int}-2+1/\lambda_1\tau_\text{eff}} \label{Delta1_Sec2}
\end{eqnarray}
and
\begin{eqnarray}
    \tau_W=\frac{\lambda_1\tau_\text{int}-1}{\lambda_1-1/\tau_\text{eff}}. \label{TW_Sec2}
\end{eqnarray}

\section{\label{primordial_black_holes}Primordial black holes}                             
Before starting this section, we need to point out the symbols in Sec.\ref{double_well} only play a descriptive role, so in the event that it does not cause confusion, we will repeat those in next sections.

\subsection{The solution to stochastic differential equation}

Now simplify, for convenience, Eq.(\ref{double_well_potential}) as
\begin{eqnarray}
    V[\phi]=\frac{1}{2}(T^2-T_C^2)\phi^2+\frac{1}{4}\lambda\phi^4+V_0, \label{double_well_potential_simplified}
\end{eqnarray}
where we have absorbed the coefficient $D$ into temperature $T$ and have considered the term with $\phi^3$ as a perturbation which will be discussed in Sec. \ref{perturbation}. Inflationary scenario predicts the inflaton rolls slowly to the bottom of a potential, so we can further simplify potential (\ref{double_well_potential_simplified}) as  \cite{PhysRevD.58.083510,KRAMERS1940284}
\begin{eqnarray}
    V[\phi]=V_0-\frac{1}{2}(T_C^2-T^2)\phi^2. \label{double_well_potential_simplified_1}
\end{eqnarray}
This expression is well approximated for the condition that a particle distributes far from the well bottom, especially for $-0.7\phi_\text{min}<\phi<0.7\phi_\text{min}$, where $\phi_\text{min}>0$ denotes the coordinate of well bottom. Thus, the Einstein field equation and equation of motion of unperturbed inflation field $\phi(t)$ with potential (\ref{double_well_potential_simplified_1}) within finite-temperature read
\begin{eqnarray}
    H_0^2\simeq -\frac{8\pi}{3M_p^2}V_0, \label{Hubble_parameter}
\end{eqnarray}
and
\begin{eqnarray}
    (3H_0+\Gamma)\dot{\phi}+V'=\xi(t). \label{EOM_unperturbed_inflaton_fluctuation}
\end{eqnarray}
The dissipation-fluctuation relation of stochastic noise $\xi(t)$ is given by
\begin{eqnarray}
    \langle\xi(t)\xi^*(t')\rangle=2 k_\text{B}T\Gamma H_0^3\delta(t-t'), \label{dissipative_fluctuation_t}
\end{eqnarray}
where $\langle\cdots\rangle$ represents the ensemble average. The solution to Eq.(\ref{EOM_unperturbed_inflaton_fluctuation}) in absence of fluctuational force $\xi(t)$ writes
\begin{eqnarray}
    \phi_\text{cl}(t)=\phi_s \text{e}^{-\frac{m^2}{3(1+r)}H_0t}, \label{solution_unpert_cl}
\end{eqnarray}
where $\phi_s$ is the initial condition, $r$ denotes the ratio between the dissipative coefficient and Hubble parameter $H_0$, i.e. $r=\Gamma/3H_0$, and
\begin{eqnarray}
    m^2={Dk_\text{B}^2(T^2_C-T^2)}/{H_0^2}. \label{m2_Sec3}
\end{eqnarray}
According to the discussions in previous section, the static probability distribution function of Eq.(\ref{EOM_unperturbed_inflaton_fluctuation}) is
\begin{eqnarray}
    W_0(\phi)=Z^{-1}\exp{\left[-A\phi^2-B\phi^4\right]}, \label{W0_phi_t}
\end{eqnarray}
with $A=-{m^2(1+r)}/{2k_\text{B}TH_0r}$ and $B={\lambda(1+r)}/{4k_\text{B}TH^3_0r}$.

If the spacial fluctuation taken into account, the Langevin equation of perturbed inflation field is given by
\begin{eqnarray}
    \frac{\partial^2}{\partial t^2}\phi(\textbf{x},t)+[3H+\Gamma]\frac{\partial}{\partial t}\phi(\textbf{x},t)&-&\frac{1}{a^2(t)}\nabla^2\phi(\textbf{x},t)\nonumber \\
    -m^2H_0^2\phi(\textbf{x},t)+\lambda\phi^3(\textbf{x},t) &=& \xi(\textbf{x},t), \label{EOM_pertubed_inflaton_PHB}
\end{eqnarray}
with $a(t)\simeq a_0\text{e}^{H_0t}$ and
\begin{eqnarray}
    \langle \xi(\textbf{x},t)\xi^*(\textbf{x},t')\rangle=2\Gamma k_\text{B} T a^{-3}\delta^3(\textbf{x}-\textbf{x}')\delta(t-t'). \label{dis_flu_relation_location_PHB}
\end{eqnarray}
By employing the physical coordinate under the coordinate transformation $\text{d}\textbf{x}_p\equiv a(t)\text{d}\textbf{x}$, the relation (\ref{dis_flu_relation_location_PHB}) becomes
\begin{eqnarray}
    \langle \xi(\textbf{x}_\text{p},t) \xi^*(\textbf{x}_\text{p},t')\rangle=2\Gamma k_\text{B} T\delta^3(\textbf{x}_\text{p}-\textbf{x}'_\text{p})\delta(t-t'). \label{dis_flu_relation_phy_PHB}
\end{eqnarray}
The Fourier transform of Eq.(\ref{EOM_pertubed_inflaton_PHB}) is approximated \cite{duderstadt1979transport,xu2017representations}
\begin{eqnarray}
    [3H_0+\Gamma]\frac{\partial}{\partial t}\tilde{\phi}(\textbf{z},t)&+&(z^2-m^2-\lambda d)H_0^2\tilde{\phi}(\textbf{z},t)\nonumber \\
    &+&\lambda'd^3H_0^6\tilde{\phi}^3(\textbf{z},t) = \xi(\textbf{z},t), \label{SDE_pertubed_inflaton_k_PHB}
\end{eqnarray}
where $\textbf{z}=\textbf{k}_\text{p}/H_0$. $\Gamma$ and $\xi(\textbf{k},t)$ appearing in Eq.~(\ref{SDE_pertubed_inflaton_k_PHB}), of course, follow the dissipation-fluctuation relation as well
\begin{eqnarray}
    &\langle& \xi(\textbf{z},t)\xi^*(\textbf{z}',t')\rangle
    =2(2\pi)^3 k_\text{B} T \Gamma H_0^{-3}\delta^3(\textbf{z}-\textbf{z}')\delta(t-t') \nonumber\\
    &=&2(2\pi)^3 k_\text{B} T H_0^{-3}(3H_0+\Gamma)\frac{r}{1+r}\delta^3(\textbf{z}-\textbf{z}')\delta(t-t'). \nonumber\\ \label{dis_flu_k_PHB}
\end{eqnarray}
The constant $d$ in Eq.(\ref{SDE_pertubed_inflaton_k_PHB}) is the variance of Eq.~(\ref{EOM_unperturbed_inflaton_fluctuation}) in terms of equilibrium probability distribution function $W_0$ obtained in Eq.~(\ref{W0_phi_t}):
\begin{eqnarray}
    d=\frac{\langle\phi^2\rangle_0}{H^2_0}=\frac{1}{\sqrt{8\pi B}H_0^2}\frac{D_{-\frac{3}{2}}(x)}{D_{-\frac{1}{2}}(x)}, \label{d_Sec3}
\end{eqnarray}
where we have used the relation of Eq.(\ref{variance_Sec2}). $\lambda'$ in Eq.~(\ref{EOM_unperturbed_inflaton_fluctuation}) is approximately evaluated as $\lambda'\simeq \lambda/2d^3$, but the value of $\lambda'$ has no significant compact on the final result.

It's noticed that the shape of potential is dependent on scale factor $z$. If $z^2<m^2+\lambda d$, $V[\phi]$ is a potential with double-well. While if $z^2>m^2+\lambda d$, $V[\tilde{\phi}]$ represents roughly a single quadratic potential. It's obvious that the scale parameter $z$ and $\lambda d$ play the role to modify the potential $V[\tilde{\phi}]$ in Eq.~(\ref{EOM_pertubed_inflaton_PHB}), so the potential in Eq.~(\ref{SDE_pertubed_inflaton_k_PHB}) may be called the effective potential, which leads to dramatically different statistical properties for different conditions with $z^2$ and $m^2+\lambda d$.

We first focus on the condition with $z^2<m^2+\lambda d$. The solution to Eq.~(\ref{SDE_pertubed_inflaton_k_PHB}) in absence of $\xi(\textbf{z},t)$ and $\lambda'd^3H_0^6\tilde{\phi}^3(\textbf{z},t)$ reads
\begin{eqnarray}
    \tilde{\phi}_\text{cl}(\textbf{z},t)&=&\phi_s H_0^{-3} \exp{\left[-\frac{\tilde{z}^2}{3(1+r)}H_0t\right]} \nonumber\\
    &=&\phi_s H_0^{-3} \exp{\left[-\frac{z^2-m^2-\lambda d}{3(1+r)}H_0t\right]} \nonumber\\
    &=&H_0^{-3} \exp{\left[-\frac{z^2-\lambda d}{3(1+r)}H_0t\right]}\phi_\text{cl}(t), \label{solution_perturbed_cl}
\end{eqnarray}
with
\begin{eqnarray}
    \tilde{z}^2=z^2-m^2-\lambda d. \label{ztilde2_Sec3}
\end{eqnarray}

According to the conclusions in Sec.\ref{double_well}, we obtain the solution to Eq.(\ref{SDE_pertubed_inflaton_k_PHB})
\begin{eqnarray}
    P(\tilde{\phi(\textbf{z},t)|\phi_s,t_0})=\frac{1}{\sqrt{2\pi} \sigma(t)}\exp{\left[-\frac{(\hat{\tilde{\phi}}(\textbf{z},t)-\tilde{\phi}_\text{cl}(\textbf{z},t))^2}{2\sigma^2(t)}\right]}, \nonumber\\ \label{conditioned_solution}
\end{eqnarray}
where $\hat{\tilde{\phi}}(\textbf{z},t)$ is a stochastic variable described by Eq.(\ref{SDE_pertubed_inflaton_k_PHB}), and $\sigma^2(t)$ is employed by
\begin{eqnarray}
    \sigma^2(t)=C(0)-C(t) \label{sigma2}
\end{eqnarray}
The autocorrelation function, of course, is the most important function we need to solve
\begin{eqnarray}
    C(t)=\tilde{d}\left[\Delta_1\text{e}^{-\lambda_1t}+(1-\Delta_1)\text{e}^{-t/\tau_W}\right], \label{autocorrelation_Sec3}
\end{eqnarray}
where $\tilde{d}$ is the variance of Eq.~(\ref{SDE_pertubed_inflaton_k_PHB})
\begin{eqnarray}
    \tilde{d}=\langle\tilde{\phi}^2\rangle_0=C(0)
    =\frac{1}{2\sqrt{2\tilde{B}}}\frac{D_{-\frac{3}{2}}(\tilde{x})}{D_{-\frac{1}{2}}(\tilde{x})}. \label{dtilde_Sec3}
\end{eqnarray}
The exact expression to $\tilde{d}$ in Eq.~(\ref{dtilde_Sec3}) is given in Sec.~\ref{normalixation}, together with $\lambda_1$ and $\tau_W$ introduced in Sec.~\ref{double_well}, so there is no need to write them again. Note that the ensemble average $\langle\cdots\rangle_0$ denotes the statistics on equilibrium PDF obtained from Eq.~(\ref{SDE_pertubed_inflaton_k_PHB}) instead of Eq.~(\ref{W0_phi_t}).

The linear perturbation, however, is no longer applied because PHBs relate the nonlinear dependence of the metric perturbation. Following the previous references \cite{Creminelli_2004,PhysRevD.42.3936}, we write the metric in the quasi-isotropic form
\begin{eqnarray}
    \text{d}s^2=-\text{d}t^2+\hat{a}^2(\hat{\phi}(\textbf{x},t))\text{d}\textbf{x}^2, \label{nonlinear_metric}
\end{eqnarray}
where scale factor $\hat{a}$ depends on spatial coordinates as well. the metric perturbation is quantified in term of
\begin{eqnarray}
    \hat{h}=\frac{\hat{a}(\textbf{x},t)}{a(t)}-1. \label{h_Sec3}
\end{eqnarray}
In the limit of small $\hat{h}\ll 1$, the metric \ref{nonlinear_metric} reduces to the standard adiabatic perturbation, which means $\hat{h}\ll 1$ leads to the gauge-invariant quantities \cite{10.1143/PTPS.78.1}. $\hat{h}$ and e-folding number $N_\text{cl}$ are related by
\begin{eqnarray}
    \hat{h}=\exp{(\hat{N}-N_\text{cl})}-1, \label{h_N_Sec3}
\end{eqnarray}
where $N_\text{cl}(t)=H_0t$ is obtained from Eq.~(\ref{EOM_unperturbed_inflaton_fluctuation}) in absence of fluctuation term $\xi(t)$ while $\hat{N}$ is a stochastic variable described by Eq.~(\ref{EOM_unperturbed_inflaton_fluctuation}) in presence of $\xi(t)$.

It's necessary to derive the expression of conditioned PDF as a function of gauge-invariance variable $h$ introduced in Eq.~(\ref{h_Sec3}). We find conditioned probability distribution funbction $P(\hat{\tilde{\phi}}(\textbf{z},t)|\phi_s)$ as
\begin{eqnarray}
    & &P(\hat{\phi}(t)|\phi_s,t_0)\nonumber\\
    &=&\frac{1}{\sqrt{2\pi} \sigma(t)}\exp{\left[-\frac{\left(\hat{\tilde{\phi}}(\textbf{z},\hat{\phi}(t))-
    \tilde{\phi}_\text{cl}(\textbf{z},\phi_\text{cl}(t))\right)^2}{2\sigma^2(t)}\right]}\frac{\text{d}\tilde{\phi}_\text{cl}}{\text{d}{\phi}_\text{cl}} \nonumber\\
    &=&\frac{1}{\sqrt{2\pi} \sigma(t)}\exp{\left[-\frac{H_0^{-6}\text{e}^{-\frac{2(z^2-\lambda d)}{3(1+r)}H_0t}}{2\sigma^2(t)}\left(\hat{\phi}(t)-\phi_\text{cl}(t)\right)^2\right]}\nonumber\\
    & &\quad\times H_0^{-3}\exp{\left[-\frac{z^2-\lambda d}{3(1+r)}H_0t\right]}. \label{conditionedPDF_phi}
\end{eqnarray}
On the other hand,
\begin{eqnarray}
    & &\hat{\phi}(t)-\phi_\text{cl}(t) \nonumber\\
    &=&\phi_s\left(\exp{\left[\frac{m^2}{3(1+r)}\hat{N}\right]}-\exp{\left[\frac{m^2}{3(1+r)}N_\text{cl}\right]}\right)\nonumber\\
    &=&\phi_\text{cl}(t)\left(\exp{\left[\frac{m^2}{3(1+r)}(\hat{N}-N_\text{cl})\right]}-1\right) \nonumber\\
    &=&\phi_\text{cl}(t)\left(\exp{\left[\frac{m^2}{3(1+r)}\ln(\hat{h}+1)\right]}-1\right)\nonumber\\
    &\simeq&\phi_\text{cl}(t)\frac{m^2}{3(1+r)}\ln(\hat{h}+1), \label{phi_h}
\end{eqnarray}
since $m^2\ll 1$. Substituting Eq.~(\ref{phi_h}) into Eq.~(\ref{conditionedPDF_phi}), we have
\begin{eqnarray}
    & &P(\hat{h}|\phi_s)=P(\hat{\phi}(\hat{h})|\phi_s)\frac{\text{d}{\hat{\phi}}}{\text{d}{\hat{h}}}\nonumber\\
    &=&\frac{1}{\sqrt{2\pi} \tilde{\sigma}(t)}\frac{1}{1+\hat{h}}\exp{\left[-\frac{\ln^2(1+\hat{h})}{2\tilde{\sigma}^2(t)}\right]}, \label{conditionedPDF_h}
\end{eqnarray}
where we have used the relation $\text{d}{\hat{\phi}}/\text{d}{\hat{N}}=\text{d}{\phi}_\text{cl}/\text{d}{N}_\text{cl}$, and
\begin{eqnarray}
    \tilde{\sigma}(t)=D^\frac{1}{2}\bar{\sigma}(t) \label{sigmatilde}
\end{eqnarray}
with
\begin{eqnarray}
    D=\tilde{d}\left(H_0^{-3}\frac{m^2}{3(1+r)}\phi_\text{s}\right)^{-2} \label{D_Sec3}
\end{eqnarray}
and
\begin{eqnarray}
    \bar{\sigma}(t)=\left(1-\frac{C(t)}{C(0)}\right)\times\exp{\left[\frac{z^2-m^2-\lambda d}{3(1+r)}H_0t\right]}. \label{sigmabar}
\end{eqnarray}

Then we focus on the condition with $z^2>m^2+\lambda d$. In this case the modified potential could be treated as a quadratic potential directly since the inflaton $\tilde{\phi}(\textbf{z},t)$ only oscillates at the bottom of potential. With this simplified condition, the autocorrelation function is given by \cite{PhysRevLett.74.1912,BERERA2000666}
\begin{eqnarray}
    C(t)=\frac{(2\pi)^2k_\text{B}Tr}{H_0^5\tilde{z}^2(1+r)}\cdot\exp{\left[-\frac{\tilde{z}^2}{3(1+r)}H_0t\right]}. \label{autocorrelation_quadratic}
\end{eqnarray}
The modified variance Eq.~(\ref{sigmatilde})is obtained
\begin{eqnarray}
    D=\frac{(2\pi)^2k_\text{B}Tr}{H_0^5\tilde{z}^2(1+r)}\cdot\left(H_0^{-3}\frac{m^2}{3(1+r)}\phi_\text{s}\right)^{-2}, \label{D_quadratic_Sec3}
\end{eqnarray}
and
\begin{eqnarray}
    \bar{\sigma}(t)=\left(1-\frac{C(t)}{C(0)}\right)\times\exp{\left[\frac{z^2-m^2-\lambda d}{3(1+r)}H_0t\right]}, \label{sigmabar_quandratic}
\end{eqnarray}
for $z^2>m^2+\lambda d$.

\subsection{\label{Abundance_PHB}Abundance of primordial black holes}
Generally, there is a threshold $h_\text{th}$ for nonlinear perturbation on formation of primordial black holes. Using Eq.~(\ref{conditionedPDF_h}), we can identify the abundance of primordial black holes
\begin{eqnarray}
    \beta(M)&=&P[h>h_\text{th}] \nonumber\\
    &=&\int_{h_\text{th}}^\infty{\frac{1}{\sqrt{2\pi} \tilde{\sigma}(t)}\frac{2}{1+\hat{h}}
    \exp{\left[-\frac{\ln^2(1+\hat{h})}{2\tilde{\sigma}^2(t)}\right]}\text{d}\hat{h}} \nonumber\\
    &=&\text{erfc}\left(\frac{\ln(1+h_\text{th})}{\bar{\sigma}(t)}\right)\nonumber\\
    &\simeq&\frac{\sqrt{2\tilde{\sigma}(t)}}{\sqrt{\pi}h_\text{th}}\exp{\left[-\frac{h_\text{th}^2}{2\tilde{\sigma}^2(t)}\right]}, \label{Abundance}
\end{eqnarray}
where $\text{erfc}(x)$ is the complementary error function. This expression, however, shows a large error on large $z$, i.e. primordial black holes with extremely small mass. This is not reasonable because the black holes with small mass exhibit an instability according to the theory of black hole radiation theory. A viable approach is to absorb the term $\exp{(-\alpha z^3 t)}$ into variance $\tilde{\sigma}(t)$. Thus the variance (\ref{sigmabar}) becomes
\begin{eqnarray}
    \bar{\sigma}(t)&=&\left(1-\frac{C(t)}{C(0)}\right)\nonumber\\
    & &\times\exp{\left[\frac{z^2-m^2-\lambda d}{3(1+r)}H_0t\right]}\cdot\exp{(-\alpha z^3 t)}. \label{sigmabar_modified}
\end{eqnarray}

\section{\label{Numerical}Numerical analysis}
The relevant formulas in previous sections are still unhelpful to get the final expressions we need. The necessary step is to normalize the formulas with
Planck mass $M_p$ or Hubble parameter $H_0$.
\subsection{\label{normalixation}Normalization}
For later purposes, it is very convenient to use proper normalized parameters or variables. By normalizing the parameters and variables
\begin{eqnarray}
    \frac{H_0^2}{M_p^2}=\frac{8\pi V_0}{3M_P^4}=\frac{8\pi\tilde{V}_0}{3}, \label{H02}
\end{eqnarray}
\begin{eqnarray}
    \bar{r}=\frac{k_\text{B}T}{M_p}, \quad \bar{r}_C=\frac{k_\text{B}T_C}{M_p}, \label{rbarC}
\end{eqnarray}
\begin{eqnarray}
    m^2=\frac{Dk^2_\text{B}(T^2_C-T^2)}{H_0^2}=\frac{3D}{8\pi}\frac{\bar{r}^2_C-\bar{r}^2}{\tilde{V}_0}, \label{m2}
\end{eqnarray}
\begin{eqnarray}
    A=-\frac{m^2(1+r)}{2k_\text{B}TH_0r}=-D\frac{\bar{r}^2_C-\bar{r}^2}{2M_p^2}\frac{1+r}{r}\left(\frac{3}{8\pi\tilde{V}_0}\right)^\frac{3}{2}, \label{A}
\end{eqnarray}
\begin{eqnarray}
    B=\frac{\lambda(1+r)}{4k_\text{B}TH^3_0r}=\frac{\lambda(1+r)}{4M_p^4\bar{r}r}\left(\frac{3}{8\pi\tilde{V}_0}\right)^\frac{3}{2}, \label{B}
\end{eqnarray}
\begin{eqnarray}
    x=\frac{A}{\sqrt{2B}}=-D(\bar{r}^2_C-\bar{r}^2)\left(\frac{1+r}{2\lambda\bar{r}r}\right)^\frac{1}{2}\left(\frac{3}{8\pi\tilde{V}_0}\right)^\frac{1}{4}, \label{x}
\end{eqnarray}
\begin{eqnarray}
    d=\frac{\langle\phi^2\rangle_0}{H^2_0}=\left(\frac{2\pi\lambda(1+r)}{\bar{r}r}\right)^{-\frac{1}{2}}\left(\frac{8\pi\tilde{V}_0}{3}\right)^{-\frac{1}{4}}
    \frac{D_{-\frac{3}{2}}(x)}{D_{-\frac{1}{2}}(x)}, \label{d}
\end{eqnarray}
\begin{eqnarray}
    \tilde{z}^2=z^2-m^2-\lambda d, \label{ztilde2}
\end{eqnarray}
\begin{eqnarray}
    \tilde{A}=\frac{\tilde{z}^2(1+r)}{2(2\pi)^3k_\text{B}TH_0^{-5}}=\frac{M_p^4\tilde{z}^2(1+r)}{2(2\pi)^3\bar{r}r}
    \left(\frac{3}{8\pi\tilde{V}_0}\right)^\frac{5}{2},  \label{Atilde}
\end{eqnarray}
\begin{eqnarray}
    \tilde{B}=\frac{\lambda d^3H_0^6(1+r)}{4(2\pi)^2k_\text{B}TH_0^{-3}r}=\frac{M_p^8\lambda d^3(1+r)}{4(2\pi)^3\bar{r}r}\left(\frac{8\pi\tilde{V}_0}{3}\right)^\frac{9}{2}, \label{Btilde}
\end{eqnarray}
\begin{eqnarray}
    \tilde{x}=\frac{\tilde{A}}{\sqrt{2\tilde{B}}}=\tilde{z}^2\left(\frac{(2\pi)^3(1+r)}{2\bar{r}r\lambda d^3}\right)^{\frac{1}{2}}
    \left(\frac{3}{8\pi\tilde{V}_0}\right)^\frac{1}{4},\label{xtilde}
\end{eqnarray}
\begin{eqnarray}
    \tilde{d}&=&\frac{1}{2\sqrt{2\tilde{B}}}\frac{D_{-\frac{3}{2}}(\tilde{x})}{D_{-\frac{1}{2}}(\tilde{x})} \nonumber\\
    &=&M_p^{-4}\left(\frac{(2\pi)^3\bar{r}r}{2\lambda d^3(1+r)}\right)^\frac{1}{2}\left(\frac{3}{8\pi\tilde{V}_0}\right)^\frac{9}{4}\frac{D_{-\frac{3}{2}}(\tilde{x})}{D_{-\frac{1}{2}}(\tilde{x})}, \label{dtilde}
\end{eqnarray}
\begin{eqnarray}
    & &D=\tilde{d}\left(H_0^{-3}\frac{m^2}{3(1+r)}\phi_\text{s}\right)^{-2}=\left(\frac{(2\pi)^3\bar{r}r}{2\lambda d^3(1+r)}\right)^\frac{1}{2} \nonumber\\
    & &\times\left(\frac{8\pi\tilde{V}_0}{3}\right)^\frac{3}{4}\left(\frac{M_p}{\phi_\text{s}}\right)^2\left(\frac{3(1+r)}{m^2}\right)^2, \;\text{for}\;\;\tilde{z}^2<0
    \label{D}
\end{eqnarray}
\begin{eqnarray}
    \tau_0 H_0&=&H_0\frac{3H_0+\Gamma}{\sqrt{2(2\pi)^3\lambda d^3H_0^6k_\text{B}TH_0^{-3}r/(1+r)}} \nonumber\\
    &=&3\left[\left(\frac{1}{\lambda \bar{r}r}\right)\left(\frac{1+r}{2\pi d}\right)^3
    \left(\frac{8\pi\tilde{V}_0}{3}\right)^\frac{1}{2}\right]^\frac{1}{2}, \label{tau0H0}
\end{eqnarray}
and
\begin{eqnarray}
    \frac{r_K}{H_0}&=&\frac{\sqrt{\big|V_\text{eff}''[0]V_\text{eff}''[\tilde{\phi}_\text{min}]\big|}}{2\pi H_0(3H_0+\Gamma)} \nonumber\\
    & &\times\exp{\left[-\frac{V_\text{eff}''[0]-V_\text{eff}''[\tilde{\phi}_\text{min}]}{(2\pi)^3k_\text{B}TH_0^{-3}r/(1+r)}\right]} \nonumber\\
    &=&\frac{\sqrt{-2\tilde{x}}}{6\pi(1+r)}\exp{\left[-\tilde{x}^2/2\right]}. \label{rKH0}
\end{eqnarray}
Based on the observations like Planck \cite{Ade:2015xua}, the parameters are evaluated as follow:
\begin{eqnarray}
    & &H_0^2/M_p^2\simeq 10^{-10},\; \lambda\simeq 10^{-14}, \;D = 0.169,\nonumber\\
    & &D\bar{r}^2\simeq 10^{-12},\; D\bar{r}_C^2\simeq 2\times10^{-12},\;r=0.5,\nonumber\\
    & &\alpha/H_0=0.075,\;\phi_s=0.05M_p,\;H_0 t_\text{end}\simeq 60,\label{parameters_evaluation}
\end{eqnarray}
where $t_\text{end}$ denotes the time at the moment of ending inflation.

\subsection{\label{numerical_result}Numerical results}
\begin{figure}
  \centering
  \includegraphics[width=3.0in,height=2.0in]{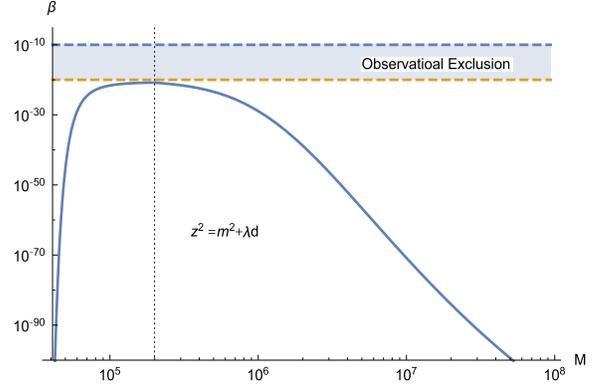}
  \caption{Expected mass spectra of primordial black holes on thermal effect in Higgs minimum model with parameters evaluated in Eq. (\ref{parameters_evaluation}). The mass is normalized in the unit the solar mass.}\label{Abundance_M}
\end{figure}

From the discussions above, we find all the information is contained into the variable $\tilde{\sigma}(t)$. The numerical analysis shows the variance is well estimated by
\begin{eqnarray}
    \tilde{\sigma}(z;t_\text{end})\simeq
    \begin{cases}
        0.009\text{e}^z+0.0123z^4,\quad z<z_c\\
        \frac{0.224}{\sqrt{2\pi}c}\exp{\left[-\frac{(z-z_c)^2}{2c^2}\right]},\;z>z_c,
    \end{cases} \label{sigmatilde_estimated}
\end{eqnarray}
with $c=0.55$ and $z_c=\sqrt{m^2+\lambda d}\simeq\sqrt{1.78}$ denoting the critical scale factor. There is an interesting property of $\tilde{\sigma}(z;t_\text{end})$ that it is insensitive to $t_\text{end}$ when scale factor distributes around the critical point $z_c$, but it quite sensitive to $t_\text{end}$ when $z$ locates far from $z_c$. Since the scale factor relates to the mass of primordial black holes, we assume the relation between mass $M$ and scale factor as
\begin{eqnarray}
    M=10^6\times M_\odot z^{-3}, \label{M_z}
\end{eqnarray}
where $M_\odot$ is the mass of solar. Instituting Eqs.~(\ref{sigmatilde_estimated}) and (\ref{M_z}) together with $h_\text{th}\simeq0.8$ \cite{PhysRevD.88.084051} into Eq.(\ref{Abundance}), we obtain the relation between abundance of PBHs $\beta(M)$ and mass of PBHs $M$ which is illustrated in Fig.~\ref{Abundance_M}. It depicts a smooth peak locating at $M$ corresponding to the critical scale factor $z_c$.

This phenomenon is not hard to explain. For the extremely large scale case, i.e. $z\rightarrow 0$, which means a deep well, the potential "force" generated by $V[\tilde{\phi}(\textbf{z},t)]$ is much stronger than the fluctuation force $\xi(\textbf{z},t)$. So the stochastic variable $\hat{\tilde{\phi}}$ moves around the classical trajectory, which shows an adiabatic perturbation under slow-roll condition. When $z^2\rightarrow m^2+\lambda d$, the potential becomes so smooth that a particle has a large probability to escape from one well into another, which describes a process that particles distribute in the well randomly. While for $z\gg 1$, particles are deeply trapped in bottom of quadratic potential.

\section{\label{perturbation}Double-well with perturbations}
In this section, we introduce two models with perturbation.

\subsection{\label{Symmetry_breaking}Symmetry breaking}
\begin{figure}
  \centering
  \includegraphics[width=3.0in,height=2.0in]{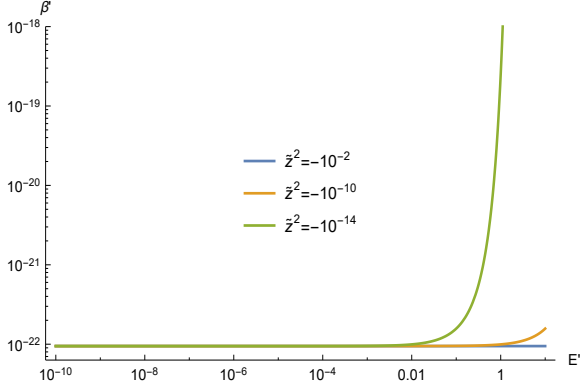}
  \caption{The difference between expected mass spectra with perturbed potential (\ref{double_well_potential}) and spectra (\ref{Abundance}) without perturbed potential. We only display the conditions with $|\tilde{z}^2|=|z^2-m^2-\lambda d|\ll 1$ since $\beta'$ almost vanishes when $|\tilde{z}^2|\sim 1$. Note, however, the shape of the large $E'$ is not exactly correct since the approximation (\ref{lambda1_modified}) is no longer hold any more.}\label{PBH_SB}
\end{figure}

As we have already mentioned in Sec.~\ref{primordial_black_holes}, we treat the term $-ET\phi^3$ as a perturbation. If this term taken into consideration, the potential is no longer symmetric on $\phi=0$. Then stochastic differential equation (\ref{SDE_pertubed_inflaton_k_PHB}) becomes
\begin{eqnarray}
    [3H_0+\Gamma]\frac{\partial}{\partial t}\tilde{\phi}(\textbf{z},t)&+&(z^2-m^2-\lambda d)H_0^2\tilde{\phi}(\textbf{z},t)\nonumber \\
    -E'H_0^4\tilde{\phi}^2(\textbf{z},t) &+&\lambda'd^3H_0^6\tilde{\phi}^3(\textbf{z},t) = \xi(\textbf{z},t). \label{SDE_pertubed_inflaton_SB_k_PHB}
\end{eqnarray}
Eq.~(\ref{SDE_pertubed_inflaton_SB_k_PHB}) may be exactly solved via continued fractions method in a manner entirely analogous  to that described by Ref. \cite{Voigtlaender1985}. For convenience, we don't attempt to solve this equation, instead, we use an easy way to deal with it. There are several parameters in autocorrelation $C(t)$ in Eq.(\ref{autocorrelation_Sec3}), but the only one needed to be modified in this case is the parameter $\lambda_1$. By expanding $V[\tilde{\phi}(\textbf{z},t)]$ near the maxima of the integrands to second order, we get \cite{risken2012fokker}
\begin{eqnarray}
    \lambda_1&=&r_{KR}+r_{KL} \nonumber\\
    &\simeq& 2r_K\left[1+\frac{H_0^2}{M_p^2}\left(\frac{|\tilde{z}^2|}{\lambda d}\right)^3\left(\frac{E'(1+r)}{3(2\pi)^3r\bar{r}}\right)^2\right]. \label{lambda1_modified}
\end{eqnarray}
In the equation above, we have assumed the perturbed term does not change the position of the potential minimum but changes its depth. Numerical result shows this perturbation will increase the probability of formation on primordial black holes at critical scale which is plotted in Fig.~\ref{PBH_SB}. This is because the perturbed term will increase the probability to escape from one well, where a particle locates, into another while decrease the probability for another well.

\subsection{\label{stochastic_resonance}Stochastic resonance}
The inflaton is often coupled with another field. Here we consider the model with Lagrangian
\begin{eqnarray}
    \mathcal{L}=\sqrt{-g}(\square\phi+\square\chi-V_\phi-V_\chi-V_I), \label{Lanranaian}
\end{eqnarray}
where $V_\chi=\frac{1}{2}m_\chi^2\chi^2$, and $V_I$ the interaction potential with
\begin{eqnarray}
    V_I=g\chi^2\phi. \label{interaction_potential}
\end{eqnarray}
We now assume field $\chi$ oscillates at its potential bottom:
\begin{eqnarray}
    \chi=A_0\exp (\text{i}m_\chi t-\gamma t), \label{chi_oscillate}
\end{eqnarray}
with damping coefficient $\gamma>0$. Thus, the stochastic differential equation reads
\begin{eqnarray}
    [3H_0+\Gamma]\frac{\partial}{\partial t}\tilde{\phi}(\textbf{z},t)&+&(z^2-m^2-\lambda d)H_0^2\tilde{\phi}(\textbf{z},t)\nonumber \\
    +g\cos(2m_\chi t)\exp(-2\gamma t) &+&\lambda'd^3H_0^6\tilde{\phi}^3(\textbf{z},t) = \xi(\textbf{z},t). \label{SDE_pertubed_inflaton_oscillate_k_PHB}
\end{eqnarray}
where we have absorbed $A_0^2$ into the interaction coefficient $g$. Eq.~(\ref{SDE_pertubed_inflaton_oscillate_k_PHB}) is deeply studied during past decades \cite{RevModPhys.70.223,PhysRevA.39.4854} which is widely used in laser optics, neuronal theory and so on, so we only give the relevant results. The drift amplitude reads
\begin{eqnarray}
    \langle\tilde{\phi}(\textbf{z},t)\rangle_0=\bar{\phi}\cos(2m_\chi t-\bar{\varphi})\exp(-2\gamma t) \label{phitilde_drift}
\end{eqnarray}
with amplitude
\begin{eqnarray}
    \bar{\phi}=\frac{g\langle\tilde{\phi}^2\rangle_0(1+r)}{(2\pi)^3k_\text{B}TH_0^{-3}r}\frac{r_K}{\sqrt{r_K^2+m_\chi^2}} \label{amplitude}
\end{eqnarray}
and phase lag
\begin{eqnarray}
    \bar{\varphi}=\arctan\left(\frac{m_\chi}{r_K}\right). \label{phase_lag}
\end{eqnarray}
Repeating the calculation in Sec. \ref{primordial_black_holes}, the conditioned probability distribution function to Eq.(\ref{SDE_pertubed_inflaton_oscillate_k_PHB}) reads
\begin{eqnarray}
    P(\hat{h};\phi_s)&=&\frac{1}{\sqrt{2\pi\tilde{\sigma}(t)}}\frac{1}{1+\hat{h}}\nonumber\\
    & &\times\exp\left[-\frac{(\ln(1+\hat{h})-\hat{D}(t))^2}{2\tilde{\sigma}^2(t)}\right],\label{conditionedPDF_SR}
\end{eqnarray}
where
\begin{eqnarray}
    & &\hat{D}(t)=\frac{g\langle\tilde{\phi}^2\rangle_0(1+r)}{(2\pi)^3k_\text{B}TH_0^{-3}r}\frac{r_K}{\sqrt{r_K^2+m_\chi^2}}
    \left(H_0^{-3}\frac{m^2}{3(1+r)}\phi_\text{s}\right)^{-1} \nonumber\\
    & &\quad\quad\times\exp\left[\frac{z^2-m^2-\lambda d-\bar{\gamma}}{3(1+r)}H_0t\right]\cdot\cos(2m_\chi t-\bar{\varphi}) \label{Dhat_Sec5}
\end{eqnarray}
with $\bar{\gamma}=6(1+r)\gamma/H_0$. Thus the abundance of PBHs in terms of Eq.(\ref{SDE_pertubed_inflaton_oscillate_k_PHB}) is given by
\begin{eqnarray}
    \beta(M)=\text{erfc}\left(\frac{\ln(1+h_\text{th})-\bar{D}(t)}{\tilde{\sigma}(t)}\right). \label{Abundance_SR}
\end{eqnarray}

It's obviously the abundance relies on the phase $2m_\chi t-\bar{\varphi}$. If the $-\pi/2<2m_\chi t_\text{end}-\bar{\varphi}<\pi/2$, the abundance decreases. While $\pi/2<2m_\chi t_\text{end}-\bar{\varphi}<3\pi/2$, it increases the probability of PBHs. This interesting result depends on both the mass of field $m_\chi$ and the barrier of the double-well potential $\tilde{x}$.

\section{\label{conclution_discussion}Conclusion and further discussion}

In this paper, we attempt a new scheme to combine the Higgs field in the minimal standard model with statistical physics together by introducing the thermal effect into the formation of primordial black holes during inflation. This scenario is something like warm inflation but is not exactly the same.

With this scenario, we find many interesting properties and results. The most important result is the exclusion to the possibility of primordial black holes with neither extremely large mass nor extremely small mass. The extremely large mass is excluded because the potential force dominates the evolution of the stochastic differential equation, while the extremely small mass is due to PBHs' instability. The peak locates at a specific coordinate of mass which corresponds to the critical scale factor $z_c^2=m^2+\lambda d$, at which the effective potential in Eq.~(\ref{SDE_pertubed_inflaton_k_PHB}) is so smooth that it may not keep stopping a particle from escaping from one well into another. This result is consistent with previous work \cite{PhysRevD.58.083510}.

Then we introduce two perturbed models, one is symmetry breaking model and another is stochastic resonance model. The former may increase the probability st the critical scale. The latter may both increase and decrease the probability due to the field mass $m_\chi$ and the barrier of effective potential $\tilde{x}$. There are many points, in fact, deserve further study for stochastic resonance model which exhibits a scenario on coupling with other fields. This model may be possible to be extended into the study on reheating or preheating \cite{PhysRevD.64.021301,PhysRevD.85.044055}.

In addition, primordial black holes in the model with double-well are studied via tunnelling probability which is a quantum process \cite{RevModPhys.57.1}. But the temperature makes it possible for a thermal effect based on which the decoherent effect should be taken into account. Such a decoherent effect is widely studied in condensation physics \cite{doi:10.1063/1.441713,PhysRevA.45.3637,Gillan_1987}. We also show a keen interest on this situation.

Besides, there are some inflationary models with periodic potential, like the natural inflation model\cite{PhysRevLett.65.3233,PhysRevD.47.426}, with potential
\begin{eqnarray}
    V_\text{np}[\phi]=\Lambda^4(1-\cos(\phi/f)).
\end{eqnarray}
This model is also deeply study in statistical physics \cite{coffey2012langevin,barone1982physics} which exhibits a possibility of Kramers' escaping. We also reckon this phenomenon corresponds to the formation of primordial black holes.

\begin{acknowledgments}
This work was supported by the National Natural Science Foundation of China (Grants No. 11575270, and No. 11235003)
\end{acknowledgments}

\end{document}